\documentclass[twocolumn]{aastex7}

\definecolor{green}{rgb}{0.15,0.7,0.2}

\definecolor{croker}{rgb}{0.82,0.596,0.894}

\definecolor{purple}{rgb}{0.56, 0.0, 1.0}

\begin{document}

\title{The CMB optical depth constrains the duration of reionization}

\author[0000-0001-9420-7384]{Christopher Cain}
\affiliation{School of Earth and Space exploration, Arizona State University, Tempe, AZ 85281, USA}
\email[show]{clcain3@asu.edu}

\author[0000-0002-3495-158X]{Alexander Van Engelen}
\affiliation{School of Earth and Space exploration, Arizona State University, Tempe, AZ 85281, USA}
\email{alexander.van.engelen@asu.edu}

\author[0000-0002-6917-0214]{Kevin~S.~Croker}
\affiliation{School of Earth and Space exploration, Arizona State University, Tempe, AZ 85281, USA}
\email{kcroker1@asu.edu}

\author[0000-0003-0238-8806]{Darby Kramer}
\email{dmkrame1@asu.edu}
\affiliation{School of Earth and Space exploration, Arizona State University, Tempe, AZ 85281, USA}

\author[]{Anson D'Aloisio}
\affiliation{Department of Physics and Astronomy, University of California, Riverside, CA 92521, USA}
\email{ansond@ucr.edu}

\author[]{Garett Lopez}
\affiliation{Department of Physics and Astronomy, University of California, Riverside, CA 92521, USA}
\email{garettld@gmail.com}

\begin{abstract}

Recently, it was pointed out that invoking a large value of the CMB optical depth, $\tau_{\rm CMB} = 0.09$, could help resolve tensions between DESI DR2 BAO data and the CMB.  This is larger than the value of $\tau_{\rm CMB} = 0.058$ measured from the Planck low-$\ell$ polarization data.  Traditionally, $\tau_{\rm CMB}$ is thought of as a constraint on reionization's midpoint. However, recent observations and modeling of the Ly$\alpha$ forest of high-$z$ quasars at $5 < z < 6$ have tightly constrained the timing of the last $10-20\%$ of reionization, adding nuance to this interpretation.  Here, we point out that fixing reionization's endpoint, in accordance with the latest Ly$\alpha$ forest constraints, renders $\tau_{\rm CMB}$ a sensitive probe of the duration of reionization, as well as its midpoint.  We compare low and high values of $\tau_{\rm CMB}$ to upper limits on the patchy kinematic Sunyaev-Zeldovich (pkSZ) effect, another CMB observable that constrains reionization's duration, and find that a value of $\tau_{\rm CMB} = 0.09$ is in $\approx 2\sigma$ tension with existing limits on the pkSZ from the South Pole Telescope.

\end{abstract}


\section{Introduction}
\label{sec:intro}
Recent measurements of baryon acoustic oscillation (BAO) signals from the Dark Energy Survey Instrument (DESI) collaboration have highlighted possible tensions in the standard $\Lambda$CDM model.   
These tensions arise when the cosmology inferred from the cosmic microwave background (CMB) as measured by \textsl{Planck} is projected forward to the redshifts probed by DESI BAO.   
They are exacerbated when including other tests of the cosmological model from moderate redshift, like CMB lensing and supernovae.  
They have been interpreted as pointing towards preferences for dynamical dark energy~\citep{DESIDR22025} or an observed excess in matter clustering, and lead to the inference of formally negative neutrino masses~\citep{Craig2024,Loverde2024,Elbers2025}.  
Together, these tensions can be viewed as arising in part from a preference by BAO data for a lower overall matter density $\Omega_{\rm m}$ than \textsl{Planck}.

Within small-scale CMB data \citet{Sailer2025} and \citet{Jhaveri2025} have highlighted the degeneracy in $\Lambda$CDM between $\Omega_{\rm m}$ and the electron scattering optical depth to the CMB $\tau_{\rm CMB}$%
\footnote{This degeneracy arises because the CMB amplitude measures the combination $A_s \exp[-2\tau_{\rm CMB}]$, where $A_s$ is the amplitude of primordial curvature fluctuations}.  
In the standard analysis, low-$\ell$ EE mode polarization from \textsl{Planck} breaks this degeneracy giving $\tau_{\rm CMB} = 0.058 \pm  0.0062$~\citep[][see also \citet{Planck2018}]{Tristram2024}.  
If one ignores the large-scale polarization measurement, the combined small-scale CMB+BAO data, including CMB lensing, prefer a somewhat higher value of $\tau_{\rm CMB} \sim 0.09$ (in $\sim 5\sigma$ tension with low-$\ell$ polarization), which reduces $\Omega_{\rm m}$ enough to significantly weaken the aforementioned $\Lambda$CDM tensions in the BAO data.  

The standard interpretation of $\tau_{\rm CMB}$ over the last decade is that it primarily contains information about {\it when} the universe reionized~\citep[e.g.][]{Zaldarriaga1997,Battaglia2013}.  The usual approach is to parameterize the reionization history with a single number%
\footnote{The current Planck data are largely limited to extracting a single parameter, namely $\tau_\mathrm{CMB}$, from their low-$\ell$ polarization measurements, while future surveys may contain further sensitivity to the ionization history~\citep[e.g.][]{Mortonson2008b}.} - its redshift midpoint $z_{\rm reion}$ - and convert $\tau_{\rm CMB}$ into a constraint on this value~\citep[e.g.][]{Komatsu2011,Planck2018}.  
Indeed, the only broadly-imposed prior on $z_{\rm reion}$ is that the Ly$\alpha$ forest of high-redshift quasars disallows a neutral fraction above $5-10\%$ at $z \lesssim 5.5$~\citep{Fan2006,McGreer2015} because of the absence of Gunn-Peterson troughs below this redshift.  

However, recent advancements in Ly$\alpha$ forest modeling and interpretation have since demonstrated that the forest is far more than a lower limit on the reionization redshift.  \citet{Becker2015} identified a $110$ $h^{-1}$Mpc Gunn-Peterson (GP) trough centered at $z = 5.7$ in the spectrum of the quasar ULAS J0148+0600, the first such trough observed at $z < 6$.  Since then, a large body of observational~\citep{Becker2018,Kulkarni2019,Zhu2021,Zhu2022,Bosman2021,Jin2023} and modeling~(\citealt{Keating2020,Keating2019,Nasir2020,Qin2021,Gaikwad2023,Cain2023,Asthana2024}) efforts have converged on a consensus that a late ($z <  6$) end to reionization is required by the Ly$\alpha$ forest data.  Further support for a late end to reionization has emerged from direct measurements of the Lyman-Continuum mean free path of the IGM from $5 < z < 6$ quasar spectra~\citep{Becker2021,Davies2021b,Cain2021,Lewis2022,Garaldi2022,Zhu2023,Satyavolu2024}.  

Most recently,~\citet{Becker2024} found strong evidence for damping wing-like features around the ULAS J0148+0600 trough, indicative of neutral gas. 
 \citet{Zhu2024} used Ly$\alpha$ similar damping wing signatures in Ly$\alpha$ forest stacks around troughs in the Ly$\beta$ forest to place the first {\it lower limits} on the IGM neutral fraction at $x_{\rm HI}(z = 5.8) \geq 6\%$ (see also~\citet{Spina2024}). 
This is consistent with other recent works that have reported constraints on the neutral fraction at $z \leq 6$ based on the scatter in Ly$\alpha$ forest opacities~\citep{Gaikwad2023,Qin2024}. 
\citet{Bosman2021} showed that large-scale fluctuations in the Ly$\alpha$ forest opacity are consistent with a fully ionized universe with a homogeneous UVB at $z < 5.3$, requiring that the IGM be not more than a few percent neutral at lower redshifts (see also the dark pixel analysis of~\citet{Jin2023}).  
Given a working definition that reionization ``ends'' when the neutral fraction is $\lesssim 5\%$\footnote{See~\citet{Gnedin2022} for a discussion of the nuances associated with defining the end of reionization.  }, these limits constrain the end of reionization to occur within the redshift range $5.3 < z < 5.8$.

Subject to this prior from the Ly$\alpha$ forest, $\tau_{\rm CMB}$ becomes more than just a constraint on reionization's midpoint.  
It becomes also a probe of reionization's duration $\Delta z$, because earlier midpoints typically require longer durations if the endpoint is constrained.
Thus, any adjustment to $\tau_{\rm CMB}$ must remain consistent with other observables that constrain $\Delta z$. 
One such observable is the Kinematic Sunyaev-Zeldovich (kSZ) effect~\citep{Sunyaev1980-pq}.  
The kSZ effect arises from the scattering of CMB photons off free electrons with kinetic velocities. Because reionization induces large-scale free electron fluctuations, it contributes significant kSZ power at tens of Mpc scales ($\ell \sim$ a few thousand), known as the ``patchy'' kSZ effect~\citep[pkSZ,][]{McQuinn2005,Park2013,Chen2022}.  
The pkSZ amplitude constrains $\Delta z$ because more power builds up the longer these ionization fluctuations persist~\citep{Zahn2005,McQuinn2005,Iliev2007,Battaglia2012}.
The pkSZ amplitude, together with higher-point functions and $\tau_{\rm CMB}$, has been suggested as a way to jointly constrain $z_{\rm reion}$ and $\Delta z$~\citep{Ferraro2018,Alvarez2021}. 
The pkSZ power spectrum amplitude has been measured at modest significance at $\ell = 3000$ with the small-scale, multi-frequency CMB maps from the South Pole Telescope (SPT)~\citep{Reichardt2020}.

The main goal of this work is to show that accounting for Ly$\alpha$ forest constraints on the timing of reionization's completion renders $\tau_{\rm CMB}$ a sensitive probe of reionization's duration $\Delta z$.  
We will also show that given these constraints, a high value of $\tau_{\rm CMB} = 0.09$ is in mild tension with recent limits on the pkSZ power from SPT.  
This work is outlined as follows.  
In \S\ref{sec:shift}, we qualitatively describe how the Ly$\alpha$ forest changes the interpretation of $\tau_{\rm CMB}$.  
In \S\ref{sec:interp}, we parameterize the reionization history and study how $\tau_{\rm CMB}$ varies in our parameter space with and without the Ly$\alpha$ forest constraint.  
In \S\ref{sec:SPTkSZ}, we study the consistency of a high value of $\tau_{\rm CMB} \sim 0.09$ with kSZ measurements from SPT by \citet{George2015} and \citet{Reichardt2020}, and we conclude in \S\ref{sec:conc}.  Throughout, we assume the following cosmological parameters: $\Omega_m = 0.3$, $\Omega_{\Lambda} = 1 - \Omega_m$, $\Omega_b = 0.045$, $h = 0.7$, $n_s = 0.96$ and $\sigma_8 = 0.8$, consistent with results from~\cite{Planck2018}. 
Distances are co-moving unless otherwise specified.
    
\section{A shift in perspective about $\tau_{\rm CMB}$}
\label{sec:shift}

\begin{figure}
    \centering
    \includegraphics[scale=0.38]{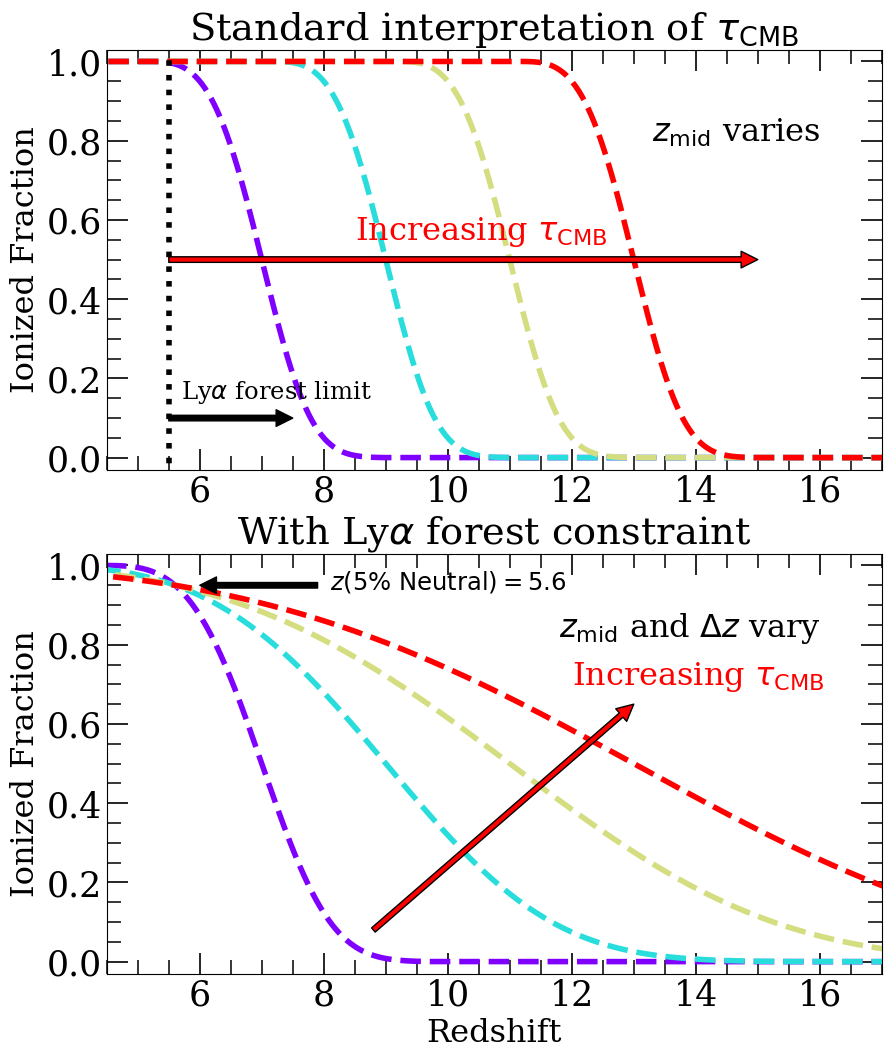}
    \caption{Qualitative illustration of how including Ly$\alpha$ forest constraints on reionization's endpoint changes the interpretation of $\tau_{\rm CMB}$.  The top panel shows several symmetric reionization histories, with higher midpoints yielding larger $\tau_{\rm CMB}$.  In this standard picture, $\tau_{\rm CMB}$ mainly constrains $z_{\rm mid}$, subject only to the Ly$\alpha$ forest requirement that reionization end before $z = 5.5$.  In the bottom panel, we pin the end of reionization ($5\%$ neutral) to $z = 5.6$, as suggested by the most recent forest data.  In this case, higher $\tau_{\rm CMB}$ requires both earlier and more extended reionization histories.  }
    \label{fig:illustration}
\end{figure}

In the top panel of Figure~\ref{fig:illustration}, we illustrate the standard approach to interpreting $\tau_{\rm CMB}$.  
We show several symmetric reionization histories with short durations, each with a different redshift midpoint, $z_{\rm mid}$.  
In this picture, a higher $\tau_{\rm CMB}$ indicates a higher $z_{\rm mid}$ and an earlier reionization.  
The unknown shape of the reionization history, including the duration, $\Delta z$, is often treated as an additional source of uncertainty in constraining the midpoint~\citep{Hu2003,Ilic2025}.  
The black vertical dotted line indicates the requirement imposed by the absence of Gunn-Peterson troughs in the Ly$\alpha$ forest much below $z \approx 6$~\citep{Fan2006}, which places a lower limit on the reionization redshift.  
Interpreted only as a lower limit on $z_{\rm mid}$, the Ly$\alpha$ forest is not very restrictive because the lowest measured $\tau_{\rm CMB}$ values from Planck suggest $z_{\rm mid} \approx 7.5-8$~\citep{Planck2018}.

When we instead leverage the Ly$\alpha$ forest as a tight constraint on the endpoint of reionization, as discussed in \S\ref{sec:intro}, this picture changes dramatically.  
In the bottom panel, we show several reionization histories with different midpoints, but all with the condition that $z(5\% \text{ neutral}) = 5.6$. 
Note that for illustrative purposes, we have assumed a symmetric reionization history, as in the top panel---we will relax that assumption below.  
Increasing $z_{\rm mid}$ (and $\tau_{\rm CMB}$) subject to this constraint requires also increasing $\Delta z$, because its endpoint is pinned to a narrow redshift range.  In this picture, $\tau_{\rm CMB}$ also contains information about {\it how long} reionization took, with both $z_{\rm mid}$ and $\Delta z$ increasing with $\tau_{\rm CMB}$.  
Thus, the Ly$\alpha$ forest imposes a key qualitative shift in perspective about the physical meaning of the CMB optical depth.

\section{Interpreting $\tau_{\rm CMB}$ in the context of the Ly$\alpha$ forest}
\label{sec:interp}

\subsection{Parameterizing the reionization history}
\label{subsec:reion}

\begin{figure*}
    \centering
    \includegraphics[scale=0.305]{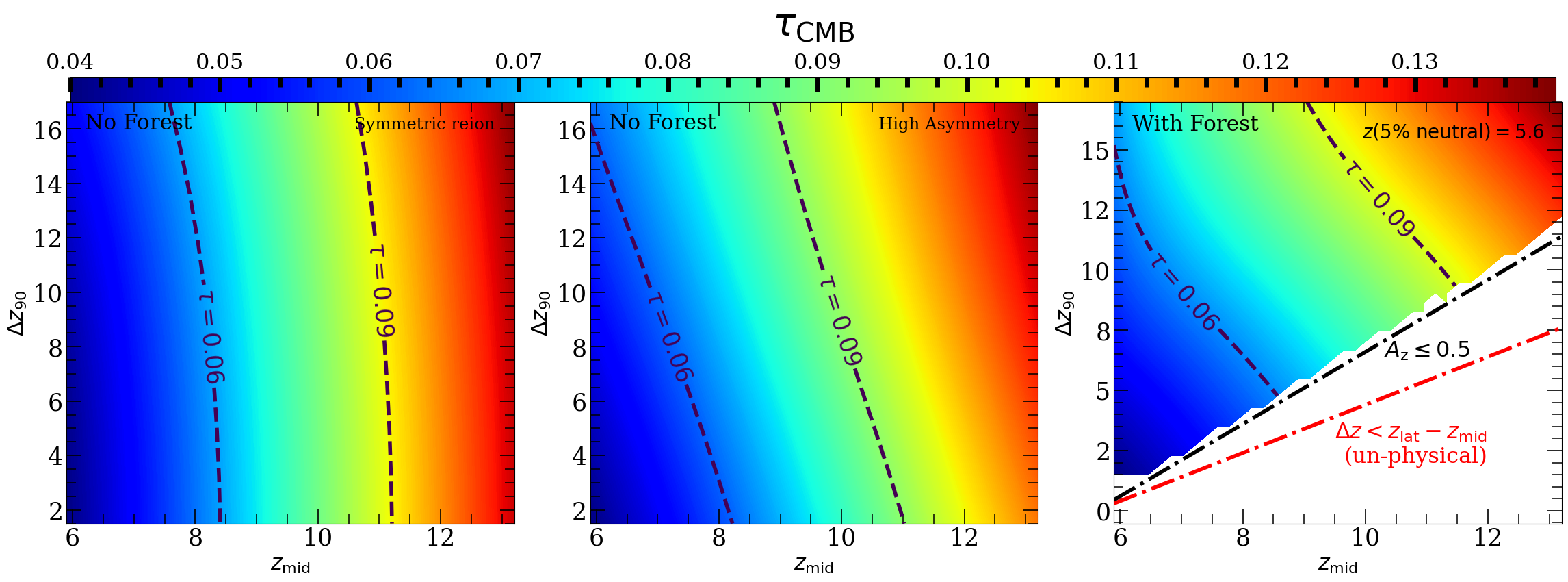}
    \caption{Dependence of $\tau_{\rm CMB}$ on $z_{\rm mid}$ and $\Delta z_{90}$.  {\bf Left:} with no Ly$\alpha$ forest constraint and assuming symmetric reionization histories ($A_z = 1$).  We show contours for $\tau_{\rm CMB} = 0.06$ and $0.09$ for reference.  {\bf Middle:} the same, but assuming a highly asymmetric reionization histories ($A_z = 5$).  {\bf Right:} including Ly$\alpha$ forest information as in Figure~\ref{fig:illustration}.  The white region in the lower right has $\Delta z < z_{\rm lat} - z_{\rm mid}$ (red line), which is un-physical, or $A_z \leq 0.5$ (black line), for which the Weibull function often does not have solutions.  In the upper left, the $\tau_{\rm CMB}$ contours are nearly diagonal, indicating sensitivity to {\it both} $z_{\rm mid}$ and $\Delta z$.  }
    \label{fig:tau_no_forest}
\end{figure*}

Throughout this work, we parameterize the reionization history following the approach outlined in~\citet{Trac2021}.  We define $z_{\rm mid}$ as the redshift at which the mass-weighted average ionized fraction reaches $50\%$.  The duration, $\Delta z$, and the ``asymmetry'', $A_{z}$, are defined by (their Eq. 10-11)
\begin{equation}
    \label{eq:reion_hist_params}
    \Delta z = z_{\rm ear} - z_{\rm lat} \hspace{1.1cm}
    A_z = \frac{z_{\rm ear} - z_{\rm lat}}{z_{\rm mid} - z_{\rm lat}}
\end{equation}
where $z_{\rm ear}$ and $z_{\rm lat}$ are the ``early'' and ``late'' reionization redshifts, respectively.  Following~\citet{Trac2021}, we take these to be the redshifts at which the ionized fraction is $5\%$ and $95\%$, respectively.  With these definitions, $\Delta z$ measures the redshift interval between $5\%$ and $95\%$ ionized, rather than the commonly adopted definition using the redshifts of $25\%$ and $75\%$ ionized~\citep[e.g.][]{Battaglia2012}.  Whenever a distinction is required between these two definitions, we will refer to them as $\Delta z_{90}$ and $\Delta z_{50}$, respectively.  Note that $A_z = 1$ corresponds to a symmetric reionization history, while for $A_z > 1$, the first half of reionization takes longer (in redshift) than the second half.  

We use the Weibull function~\citep{Weibull1951} to characterize the shape of the reionization history, as described in~\citet{Trac2021}.  As discussed in that work, the three-parameter Weibull function describes well the expected shape of reionization histories seen in analytical models and simulations, and has solutions for most physically reasonable values of the reionization parameters.  Most notably, the Weibull function can accommodate reionization histories that are highly asymmetric, with late midpoints and extended, high-redshift tails, which is important for this work.  We refer the reader to~\citet{Trac2021} for further details.

\subsection{Sensitivity of $\tau_{\rm CMB}$ to $\Delta z$}
\label{subsec:tausens}

In Figure~\ref{fig:tau_no_forest}, we show the full dependence of $\tau_{\rm CMB}$ on $z_{\rm mid}$ and $\Delta z_{90}$.
In each panel, the color-scale shows $\tau_{\rm CMB}$ as a function of $z_{\rm mid}$ and $\Delta z$ across a wide range of reionization parameter space.  The dashed lines denote iso-contours of $\tau_{\rm CMB} = 0.06$ and $0.09$ - the ``low'' and ``high'' values suggested by~\citet{Sailer2025}, the latter of which they showed can partially resolve tensions between DESI BAO data and the CMB, but which is not in agreement with the measured large-scale polarization from \textsl{Planck}.  In the left and middle panels, we do not impose any condition on $z_{\rm lat}$ from the forest, and we assume asymmetry parameters of $A_z = 1$ and $5$, respectively.  In the left panel, we see that the iso-contours are nearly vertical, indicating that $\tau_{\rm CMB}$ is almost exclusively sensitive to $z_{\rm mid}$, even for large $\Delta z$.  This is because the integral of the electron density over a symmetric reionization history is well-approximated by an integral over a delta function at the midpoint.  

Increasing the asymmetry to $A_z = 5$ introduces some sensitivity to $\Delta z$, since the free electron column density over each half of reionization is asymmetric around $z_{\rm mid}$.  However, even for large $A_z$, $\tau_{\rm CMB}$ remains more sensitive to $z_{\rm mid}$ than to $\Delta z$.  A factor of $7$ change in $\Delta z$ (from $2$ to $14$) only shifts $z_{\rm mid}$ by $\Delta z_{\rm mid} \approx 1.75$.  This shift is comparable to the $\pm 1 \sigma$ uncertainties on $z_{\rm mid}$ inferred from the latest Planck measurements of $\tau_{\rm CMB}$~\citep[Eq. 18 in][]{Planck2018}.  

In the right-most panel, we fix $z_{\rm lat} = 5.6$, as in the bottom panel of Figure~\ref{fig:illustration}.  For a given $z_{\rm mid}$ and $\Delta z_{90}$, we solve for the value of $A_z$ that yields $z_{\rm lat} = 5.6$ using Eq.~\ref{eq:reion_hist_params}.  By virtue of this condition, histories with smaller $z_{\rm mid}$ and large $\Delta z_{90}$ also have large $A_z$.    Two key features are evident: (1) the iso-contours in the upper left have rotated, and are almost diagonal across the parameter space and (2) the bottom right corner is empty.  

The first effect results from the fact that if $z_{\rm lat}$ is fixed, increasing $z_{\rm mid}$ requires an increase in $\Delta z$, as discussed in \S\ref{sec:shift}.  
The second feature arises because the Ly$\alpha$ forest constraint renders some reionization histories unlikely or even un-physical. Points in parameter space below the red dot-dashed line have $\Delta z < z_{\rm lat} - z_{\rm mid}$, and are thus un-physical.  The range between the dashed red and black lines have small asymmetry parameters $A_z \leq 0.5$ (for which the first half of reionization is twice as fast as the second), which mostly lack solutions in the parameterization described  in \S\ref{subsec:reion}.  Models with $A_z$ this small require high ionizing output from galaxies and/or AGN at early times so that the first half of reionization happens quickly, and low output at later times so that it finishes slowly.  This is unexpected given that the galaxy and AGN populations grow rapidly with decreasing redshift.  Even scenarios which rely on the faintest sources to drive extended reionization histories struggle to produce $A_z < 1$~\citep[e.g.][]{Finkelstein2019}.  

\begin{figure*}
    \centering
    \includegraphics[width=\linewidth]{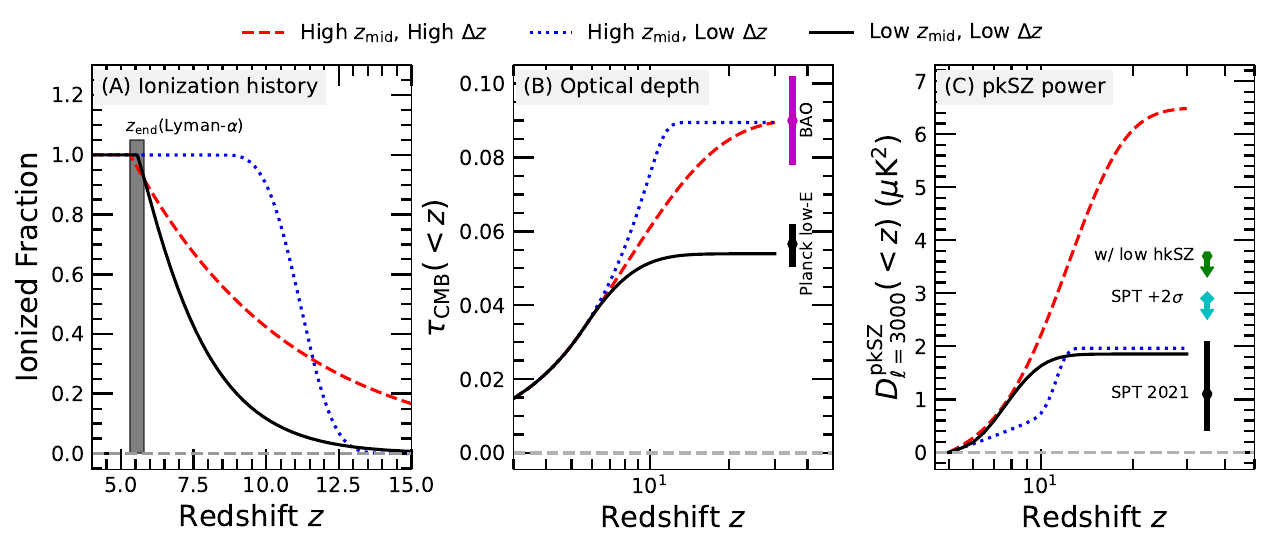}
    \caption{Representative reionization histories with different $\tau_{\rm CMB}$ and $D_{\ell=3000}^\mathrm{pkSZ}$.  From left to right, the panels show the ionization history, $\tau_{\rm CMB}(< z)$, and $D_{\ell=3000}^\mathrm{pkSZ}(< z)$ for models with low $z_{\rm mid}$/low $\Delta z$ (black solid), high $z_{\rm mid}$/high $\Delta z$ (red dashed), and high $z_{\rm mid}$/low $\Delta z$ (blue dotted).  The left panel shows the range of $z_{\rm end}$ allowed by the Ly$\alpha$ forest, the middle shows measurements of $\tau_{\rm CMB}$ (see text), and the right panel shows measurements/upper limits on $D_{\ell=3000}^\mathrm{pkSZ}$ from SPT.  The models with high $\tau_{\rm CMB}$ either end reionization too early for the Ly$\alpha$ forest (High $z_{\rm mid}$, Low $\Delta z$) or have $D_{\ell=3000}^\mathrm{pkSZ}$ too high for SPT (High $z_{\rm mid}$, High $\Delta z$).  See text for details.  }
    \label{fig:ksz_tau_examples}
\end{figure*}

\section{Relating $\tau_{\rm CMB}$ to the pkSZ}
\label{sec:SPTkSZ}

\subsection{Modeling the pkSZ}
\label{subsec:ksz}

We estimate the pkSZ signal from reionization using the AMBER code of~\citet{Trac2021}~\citep[see also][]{Chen2022}.  AMBER is an efficient semi-numerical reionization code that can rapidly compute the density, velocity, and ionization fields in sufficiently large volumes to fully capture the reionization kSZ power at $\ell = 3000$.  
In particular, AMBER calculates the reionization history by ``abundance-matching'' the fluctuations in the photo-ionization rate to the reionization redshift field, which yields a prediction for the ionization field at each redshift.  
In addition to $z_{\rm mid}$, $\Delta z$, and $A_z$, AMBER takes two additional parameters: the minimum mass of halos producing ionizing photons, $M_{\min}$, and the ionizing photon mean free path, $l_{\rm mfp}$.  Throughout, we adopt the fiducial values used in~\citet{Chen2022} of $M_{\min} = 10^8$ $M_{\odot}$ and $l_{\rm mfp} = 3$ $h^{-1}$Mpc.  Note that~\citet{Chen2022} found that the effect of these parameters is sub-dominant to that of the reionization timing parameters, introducing at most $10-15\%$ uncertainty in the pkSZ (see their Fig. 13).    

Modeling the kSZ accurately using simulations requires capturing the large-scale fluctuations in the velocity field, which necessitates box sizes of order $\sim 1$ Gpc~\citep{Park2013}.  \citet{Chen2022} found that a box size of 2 $h^{-1}$Gpc is needed to fully converge on the kSZ amplitude at $\ell = 3000$.  Exploring the reionization parameter space as required by this work involved running over $1000$ AMBER simulations.  To balance accuracy and computational efficiency, we ran our simulations in 1 $h^{-1}$Gpc using $N = 1024^3$ grid cells (matching the fiducial spatial resolution in~\citealt{Chen2022}).  
We have tested this choice against larger boxes, and found that the kSZ power is under-estimated by at most $10\%$, which is acceptable for the purposes of this work.  As we will see, this underestimation is conservative relative to our main conclusions.  

\subsection{Consistency with the SPT measurement}
\label{subsec:SPT}

\begin{figure*}
    \centering
    \includegraphics[scale=0.42]{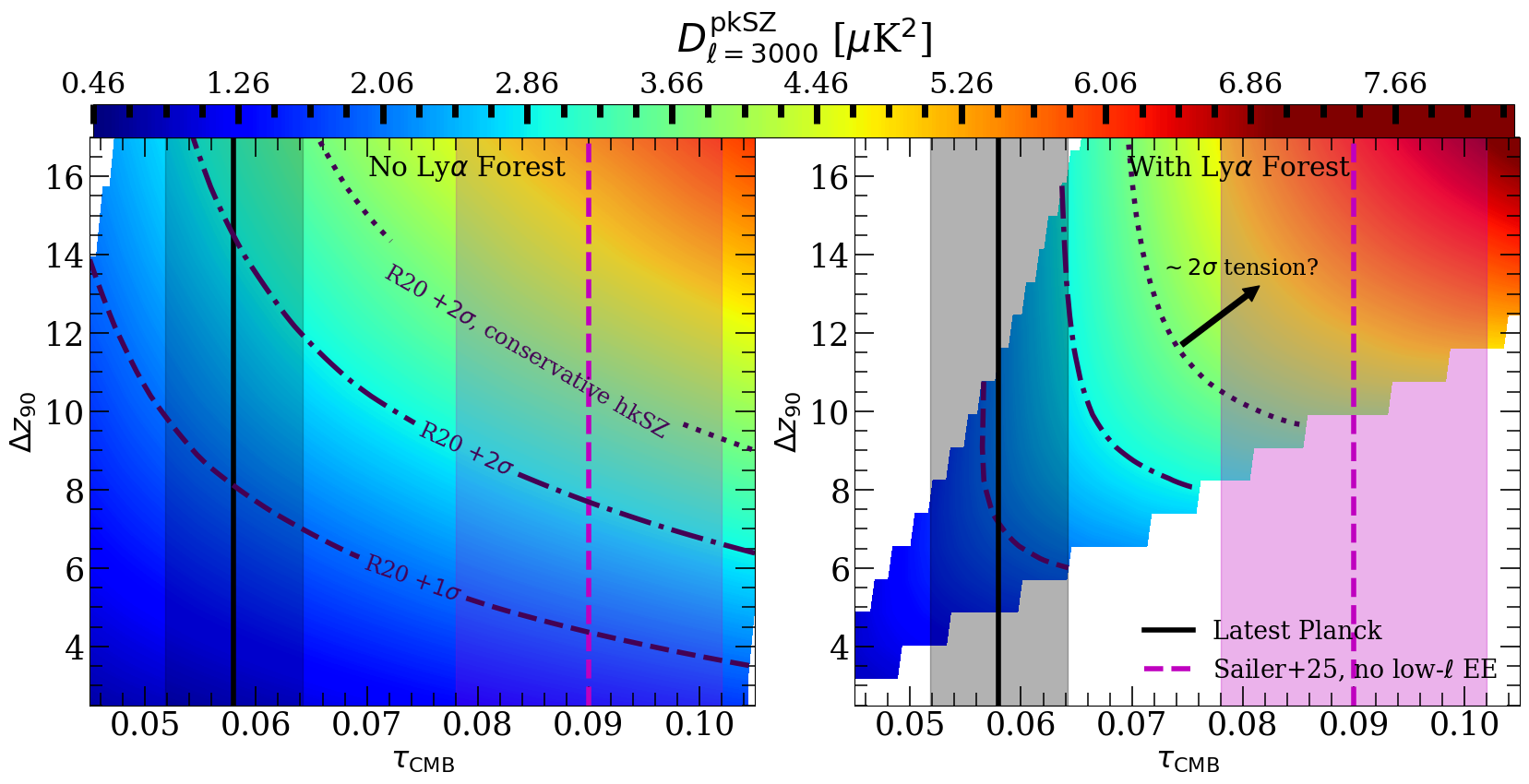}
    \caption{Consistency of upper limits on the pkSZ signal from SPT~\citep[][R20]{Reichardt2020} with low and high values of $\tau_{\rm CMB}$.  {\bf Left}: $D_{\ell=3000}^{\rm pkSZ}$ vs. $\tau_{\rm CMB}$ vs. $\Delta z_{90}$ with no Ly$\alpha$ forest constraint imposed.  The solid vertical lines indicate low and high values of $\tau_{\rm CMB}$ (see legend) and the contours denote different upper limits from~\citet{Reichardt2020} (see text).  Low values of $D_{\ell=3000}^{\rm pkSZ}$ are consistent with high $\tau_{\rm CMB}$ if reionization happened early and had a short duration (and thus ended early, lower right of the plot).  {\bf Right:} the same, but imposing the Ly$\alpha$ forest constraint.  Models with high $\tau_{\rm CMB}$ and low $\Delta z_{90}$ (and vice-versa) are excluded, and $\tau_{\rm CMB} = 0.09$, as suggested by~\citet{Sailer2025} to reconcile BAO tensions, requires $D_{\ell=3000}^{\rm pkSZ}$ above the $2\sigma$ upper limit from~\citet{Reichardt2020}.  }
    \label{fig:ksz_tau_plot_all}
\end{figure*}

Measurements of the kSZ effect from millimeter surveys on small angular scales, such as the South Pole Telescope (SPT) \citep{George2015,Reichardt2020}, the Atacama Cosmology Telescope \citep{Louis2025} and \textsl{Planck} \citep{Planck2018} place constraints on reionization.  
In particular, the constraint on the pkSZ from~\citet{Reichardt2020}, $D_{\ell=3000}^{\rm pkSZ} < 2.9$ $\mu$K$^2$, comes from combining their measurement of the {\it total} kSZ ($3.0 \pm 1.0$ $\mu$K$^2$, $1\sigma$) with a model of the post-reionization (homogeneous) kSZ power~\citep[hkSZ,][]{Shaw2012} and a measurement of the thermal Sunyaev-Zeldovich tSZ bi-spectrum~\citep{Crawford2014}.  Using these measurements, they report an upper limit on the duration of reionization of $\Delta z_{50} < 4.1$.  As we have shown in \S2-\S3, $\tau_{\rm CMB}$ is also a probe of $\Delta z$ once Ly$\alpha$ forest constraints are incorporated into the reionization history.  In this section, we use our suite of AMBER simulations to determine the implications of the recent SPT measurement of the pkSZ for $\tau_{\rm CMB}$ given these constraints.   

In Figure~\ref{fig:ksz_tau_examples}, we show several representative examples of reionization histories with different $\tau_{\rm CMB}$ and pkSZ power spectra.  From left to right, the panels show the reionization history, cumulative $\tau_{\rm CMB}$ vs. redshift, $\tau_{\rm CMB}(< z)$, and the cumulative dimensionless pkSZ power, $D_{\ell=3000}^{\rm pkSZ}(< z)$, calculated using AMBER.  The black solid curve has a relatively short duration ($\Delta z_{90} = 6$), low $\tau_{\rm CMB} = 0.054$, and a low $D_{\ell = 3000}^{\rm pkSZ} \approx 2$ $\mu$K$^2$.  The red dashed curve has a much longer duration of $\Delta z_{90} = 16$, high $\tau_{\rm CMB} = 0.09$, and large $D_{\ell = 3000}^{\rm pkSZ} \approx 7$ $\mu$K$^2$.  The blue dotted curve has a short duration of $\Delta z_{90} = 3$, a large $\tau_{\rm CMB} = 0.09$, and a small $D_{\ell = 3000}^{\rm pkSZ} \approx 2$ $\mu$K$^2$.  The thin vertical shaded region on the left panel shows the range of reionization end times allowed by the Ly$\alpha$ forest, $5.3 < z < 5.8$.  In the middle panel, we show the measured $\tau_{\rm CMB}$ from Planck low-$\ell$ EE polarization (black), and the best-fit $\tau_{\rm CMB}$ inferred from BAO data in~\citet{Sailer2025} without including Planck polarization data (magenta).  The right panel shows the measured $D_{\ell = 3000}^{\rm pkSZ} = 1.1_{-0.7}^{+1.1}$ $\mu$K$^2$ from~\citet{Reichardt2020} (black point), their fiducial $2\sigma$ upper limit ($< 2.9$ $\mu$K$^2$, cyan point), and a more conservative upper limit (green point) using recent estimates of the hkSZ from AGORA~\citep{Omori2024}, which uses gas profiles in massive halos measured from the BAHAMAS simulations~\citep{McCarthy2017}.  Specifically, their ``$\log(T_{\rm AGN}) = 8.0$'' model\footnote{This model assumes a conservatively large amount of AGN feedback in massive halos, which smears out halo gas profiles and reduces the hkSZ signal~\citep{Park2018}.  Recent observations have shown that the level assumed in this model may be realistic~\citep{Bigwood2024}.  } has their lowest estimate of $D_{\ell = 3000}^{\rm hkSZ} \approx 0.85$ $\mu$K$^2$, which is smaller than the fiducial hkSZ model assumed in~\citet{Reichardt2020}.

We see that both of the models with $\tau_{\rm CMB} = 0.09$ violate either the Ly$\alpha$ forest constraint on reionization's end (blue dotted curve), or the pkSZ limits from SPT (red dashed curve).  If reionization has a high midpoint (so that $\tau_{\rm CMB}$ is large) and a short duration (so that $D_{\ell = 3000}^{\rm pkSZ}$ is small), it will end too early for the Ly$\alpha$ forest.  The other case ends at the right time for the forest, but has a large duration and thus a $D_{\ell = 3000}^{\rm pkSZ}$ too high for the SPT measurement.  The only model that respects both constraints has a low $\tau_{\rm CMB}$, in accord with the Planck low-$\ell$ polarization measurement.  This qualitatively demonstrates the main point of this section - {\it that models with high $\tau_{\rm CMB}$ are inconsistent with either the pkSZ measurement from SPT or the Ly$\alpha$ forest data.  }

Figure~\ref{fig:ksz_tau_plot_all} shows $D_{\ell = 3000}^{\rm pkSZ}$  across our full suite of AMBER simulations.  We show $\tau_{\rm CMB}$ on the x-axis and $\Delta z_{90}$ on the y-axis~\footnote{Note that at fixed $\Delta z_{90}$, there is a one-to-one relationship between $z_{\rm mid}$ and $\tau_{\rm CMB}$ provided either $A_z$ is held fixed or the Ly$\alpha$ forest constraint is imposed (constant $z_{\rm lat}$).  }.  The vertical lines and shaded regions show the mean and $\pm 1\sigma$ intervals of $\tau_{\rm CMB}$ from~\citet{Tristram2024} (black solid) and the best-fit to BAO data, excluding low-$\ell$ EE Planck data, from~\citet{Sailer2025} (magenta dashed).  The dashed and dot-dashed iso-contours show are the fiducial $1$ and $2\sigma$ upper limits on $D_{\ell=3000}^{\rm pkSZ}$ from~\citet{Reichardt2020}.  The dotted line shows the $2\sigma$ upper limit assuming the aforementioned conservative hkSZ power estimate from AGORA (see green point in Figure~\ref{fig:ksz_tau_examples}).

The left panel shows the case with no Ly$\alpha$ forest constraint and assuming a symmetric reionization history ($A_z = 1$, as in the left-most panel of Figure~\ref{fig:tau_no_forest}).  Here, there is no problem having a simultaneously high $\tau_{\rm CMB}$ and low $D_{\ell=3000}^{\rm pkSZ}$.  Indeed, close to $\tau_{\rm CMB} = 0.09$, the iso-contours are closer to horizontal than to vertical, indicating that $\tau_{\rm CMB}$ and $D_{\ell=3000}^{\rm pkSZ}$ are mostly probing different aspects of the reionization history ($z_{\rm mid}$ and $\Delta z$, respectively).  Reionization histories with short durations, early midpoints, and early endpoints can easily satisfy the need for a high $\tau_{\rm CMB}$ and low $D_{\ell=3000}^{\rm pkSZ}$.  

However, the right panel demonstrates that imposing the Ly$\alpha$ forest constraint completely changes this picture.  The parts of parameter space with high $\tau_{\rm CMB}$ and low $\Delta z_{90}$, and vice versa, are blank, indicating that such scenarios are disallowed by the Ly$\alpha$ forest.  We also see that $\tau_{\rm CMB}$ and $D_{\ell=3000}^{\rm pkSZ}$ are now strongly correlated, since both are sensitive to $\Delta z_{90}$.  The iso-contours have also shifted, such that the part of parameter space with high $\tau_{\rm CMB}$ that is allowed by the forest is almost entirely above our most conservative $2\sigma$ upper limit on $D_{\ell=3000}^{\rm pkSZ}$ based on the total kSZ measurement from~\citet{Reichardt2020} .  The low $\tau_{\rm CMB}$ measured by Planck remains comfortably below these limits on $D_{\ell=3000}^{\rm pkSZ}$.  

Several caveats are worth mentioning.  First, the limits on $D_{\ell=3000}^{\rm pkSZ}$ from~\citet{Reichardt2020} depend not only on the hkSZ, but also on the functional form of the reionization history and models for other components of the total high-$\ell$ signal, such as the cosmic infrared background (CIB)~\citep[e.g.,][]{Viero2013,Planck2014CIB}, the tSZ~\citep[e.g.,][]{Komatsu2002,Shaw2010}, and their correlation.  While there was some exploration of the dependence of the result on these assumptions, it is possible that these much-brighter components are more complex than previously assumed.  It has also been recently shown that there may be other non-negligible sources of extragalactic emission at small angular scales, such as emission from  CO and its correlation with the CIB \citep{Maniyar2023,Kokron2024}. 

The parameterization of the reionization history assumed here also limits us to well-behaved, monotonic functional forms that have $A_z \gtrsim 0.5$, excluding more exotic reionization scenarios, such as some population III-driven scenarios~\citep{Qin2020,Wu2021c} and/or ``double reionization'' models~\citep{Cen2003,Furlanetto2005b,Ahn2021}.  Variations in the $M_{\min}$ and $l_{\rm mfp}$ parameters in AMBER and/or more accurate modeling of the ionization field~\citep[e.g., using radiative transfer simulations, ][]{Mellema2006,Trac2007,Cain2024c} could also affect this tension.  Lastly, although we have fixed $z(5\% \text{ neutral}) = 5.6$ in this analysis for simplicity, in reality there is a small  uncertainty ($\Delta z \sim 0.5$) on the Ly$\alpha$ forest constraint on reionization's end.  This could also slightly affect the strength of the tension, and should be accounted for in more formal forthcoming analyses.  

\section{Discussion \& Conclusions}
\label{sec:conc}

We have examined the implications of $\tau_{\rm CMB}$ measurements for the reionization history given tight constraints on reionization's end from the Ly$\alpha$ forest.  We find: 

\begin{enumerate}

    \item A tight constraint on reionization's end from the Ly$\alpha$ forest renders $\tau_{\rm CMB}$ a sensitive probe of reionization's duration, as well as its midpoint.

    \item Given the Ly$\alpha$ forest constraint, a high value of $\tau_{\rm CMB} = 0.09$ that relaxes tensions between BAO data and the CMB is in $\approx 2\sigma$ tension with existing upper limits on the pkSZ from the CMB~\citep{Reichardt2020}.  
    The significance of this tension is sensitive to assumptions about the shape of the reionization history and foregrounds that effect the pkSZ measurement (such as the CIB and tSZ) and will require further work to confirm.  
    
\end{enumerate}

At face value, our findings disfavor a value of $\tau_{\rm CMB} = 0.09$ (even without reference to polarization data), and thus prefer scenarios that can resolve the BAO-CMB tensions without invoking a high $\tau_{\rm CMB}$~\citep[e.g][]{Ahlen2025, ChenZaldarriaga2025, KumarAjith2025}.  
We note that the Ly$\alpha$ forest constraint will also impact inference of $\tau_{\rm CMB}$ from cosmological datasets through the assumed functional form of the reionization history~\citep{Hu2003,Mortonson2008,Mortlock2011,Jhaveri2025}.  
Although \citet{Qin2020} explore more realistic reionization histories and find little preference for a higher optical depth in \textsl{Planck} 2018 data, they do not perform cosmological inference with all $\Lambda$CDM parameters.  
While other cosmological parameters are well-constrained by the CMB spectra, the inclusion of DESI BAO shifts $\Omega_m$ at the $\sim 5\%$ level~\citep{DESIDR22025}.
The degeneracy between $\Omega_m$ and $\tau_{\rm CMB}$, given the reionization histories considered here and constraint from various CMB datasets with BAO, is the subject of future work.

Another possibility is that forthcoming observations will revise $D_{\ell=3000}^{\rm pkSZ}$ higher than measured by~\citet{Reichardt2020} (see caveat in \S\ref{subsec:SPT}), in which case a high $\tau_{\rm CMB}$ may become an allowed, if not preferred, scenario.  If the pkSZ limits are later revised higher, either due to a larger central value or wider error bars, the preference for a low $\tau_{\rm CMB}$ pointed out here may disappear.  
Forthcoming experiments, such as Simons Observatory \citep{SimonsObs2019,SimonsObs2025}, SPT-3G \citep{Benson2014}, CMB-S4~\citep{Carlstrom2019}, and CMB-HD~\citep{Sehgal2020} promise to measure the pkSZ power spectrum and infer reionization's duration and spatial morphology.  
They will also measure higher-order statistics, both on their own ~\citep[e.g.][]{Smith2017,Ferraro2018,Alvarez2021} and in conjunction with other probes \citep[][]{LaPlante2022,Kramer2025}; upper limits for these higher-point pkSZ functions have recently been obtained  \citep{Raghunathan2024,MacCrann2024}.  These measurements should constrain the aforementioned statistics to sub-percent precision, potentially exacerbating the tension with a high $\tau_{\rm CMB}$ described in this work.

\begin{acknowledgments}
The authors thank Matthew McQuinn, Miguel Morales, and Simon Mork for helpful discussions. Some computations for this work were run using computational resources distributed under NSF ACCESS allocations TG-PHY240332, and used the Bridges-2 supercomputer at the Pittsburg Supercomputing Center.  CC acknowledges support from the Beus Center for Cosmic Foundations. AD acknowledges support from
NSF grant AST-2045600.    AvE and DK acknowledge support from  NASA grants 80NSSC23K0747, 80NSSC23K0464, and 80NSSC24K0665, and NSF grant 588167.

\end{acknowledgments}

\bibliography{references}{}
\bibliographystyle{aasjournalv7}

\end{document}